\documentclass[aps,prc,12pt,amsmath,amssymb,nofootinbib,superscriptaddress,preprint]{revtex4-1}
\usepackage{selinput}
\usepackage[T1]{fontenc}
\usepackage{graphicx}
\usepackage{verbatim}
\usepackage[dvipsnames]{xcolor}
\usepackage{geometry}
\usepackage{units}
\usepackage[toc,page]{appendix}
\usepackage{float}
\usepackage{lipsum}
\usepackage{hyperref}
\usepackage{bm}
\usepackage{color}
\usepackage{ulem}

\newcommand \beq{\begin{equation}}
\newcommand \eeq{\end{equation}}

\newcommand{\Th}{T_{\mathrm{H}}}
\newcommand{\Nc}{N_{\mathrm{c}}}
\newcommand{\Nf}{N_{\mathrm{f}}}

\newcommand{\Tr}{\mathop{\mathrm{Tr}}}

\newcommand{\MeV}{\;\text{MeV}}

\newcommand{\rr[1]}{{\color{red}{#1}}}
\newcommand{\rs[1]}{{\color{red}{\bf \sout{#1}}}}

\begin{document}
\title{Two Lectures on the Phase Diagram of QCD\footnote{Two lectures presented at the Cracow School of Physics: Fundamental Interactions$-$65 Years of the Cracow School, June 14-21,
2025, Zakopane, Tatra Mountains, Poland.}}

\author{Larry McLerran}
\affiliation{Institute for Nuclear Theory, University of Washington, Box 351550, Seattle, WA 98195, USA}

\begin{abstract}
The phase diagram of QCD at finite temperature and density is discussed.  Large numbers of quark colors, $N_{\rm c} >> 1$, is used to explain generic features of the phase diagram.    For temperatures below $ T \le 160$~MeV at zero baryon number density, the three dimensional string model is shown to describe the thermodynamics of QCD, and as well, the integrated spectrum of non-Goldstone mesons and glueballs.  The lowest mass state in the spectrum of the open and closed string is treated separately due to the tachyon problem of string theory. This is with no undetermined free parameters.  It is argued that there are at least three phases at zero baryon number density  characterized by the $N_{\rm c}$ dependence of extensive thermodynamic quantities.  It is also argued that the intermediate phase has restored chiral symmetry.  At high baryon number density and low temperature, again there are three phases.  A Quarkyonic phase, with energy density of order $N_{\rm c}$,  is distinguished from its counterpart at low baryon density and temperature by its chiral properties.
\end{abstract}

\maketitle

\section{Introduction}

This talk presents a summary of results done with and by my collaborators Rob Pisarski, Yoshimasa Hidaka, Toru Kojo, Krzysztof Redlich, Chihiro Sasaki, Sanjay Reddy, 
 Saul Hernandez-Ortiz,  Kiesang Jeon, Srimoyee Sen, Michal Praszalowicz,  Yuki Fujimoto, Kenji Fukushima, Michal Marczenko,   Vlodymyr Vochenko, 
 Volker Koch, Jerry Miller, Marcus Bluhm, Marlene Nahrgang, and Gyozo Kovacs, in 
 Refs.~\cite{McLerran:2007qj,Hidaka:2008yy,McLerran:2008ua,McLerran:2018hbz,Jeong:2019lhv,Fujimoto:2022ohj,Kojo:2011cn,Kojo:2021ugu,
Fujimoto:2023mzy,Marczenko:2022jhl,Koch:2024qnz,McLerran:2024rvk,Bluhm:2024uhj,Fujimoto:2025sxx,Marczenko:2025nhj}.  
I attempt to explain these works as a coherent and modern picture of the QCD phase diagram.  
Of course I wish to make the picture compelling and as such, my presentation will by some considered to be overly enthusiastic.  So be it.  
The papers cited here are
only a small sampling of the tremendous effort put into this field, which is foundational 
for the work presented here, but my goal in these lectures is to present the ingredients as an introduction to this body of research, and not to provide 
a comprehensive summary of the work done in this area.   

The first part of these lectures will concern itself with the properties of matter at finite temperature and zero baryon number density.  The second part will focus on the properties of matter at high baryon density and low temperature.

\section{Matter at Finite Temperature and Low Baryon Number Density}

The story I tell here might have been told many years ago, but for some reason was not.  The scientific questions addressed  are
\begin{enumerate}
    \item{Are there phases of matter between the Quark Gluon Plasma and the Hadron Gas?}
    \item{Can one quantitatively describe the thermodynamics in the hadron phase, which at zero baryon density is for temperatures less than $T \le 160$~MeV?}
    \item{What is the physics beyond the hadron phase? }
\end{enumerate}
This work is in part motivated by some crucial observations of Cohen and Glozman \cite{Cohen:2023hbq}, and earlier observations of myself and Rob Pisarski~\cite{McLerran:2007qj}.

At zero baryon number density, the thermodynamics of QCD is well studied using lattice Monte Carlo methods~\cite{Bazavov:2011nk},\cite{Borsanyi:2013bia}. 
These studies demonstrate that chiral symmetry is broken until a temperature of about $T \sim ~160$~MeV.
On the other hand, the expectation value of the Polyakov line indicates confinement disappears at a higher temperature $T \sim 300$~MeV.  
This is presumably the temperature of deconfinement. It also corresponds to the temperature of deconfinement of a theory with no dynamical quarks.
In the temperature range $160~{\rm MeV} \le T \le 300~\rm{MeV}$, the number of degrees of freedom as measured by the entropy density scaled by $T^3$, 
$N_{\rm dof} \sim s/T^3$ is smaller than expected for an ideal gas of massless quarks and gluons. 
We will see that this is because in this intermediate temperature range, the degrees of freedom of gluons are locked inside confined 
glueballs, and because glueballs are quite heavy, the states do not contribute much to the entropy.

The above statements are very controversial and would be disputed by many, but in this lecture, I will argue that they are quite reasonable, and provide a quantitative framework in which they are justified.

Part of the reason for controversy  is that it is difficult to characterize deconfinement at finite temperature. 
At zero temperature and zero baryon density, confinement simply means that no observable states carry fractional charge, 
corresponding to a deconfined quark.  At finite temperature, such a characterization will not work since thermal fluctuations 
will make the isolation of a single particle state impossible.  This is because fluctuations can produce particles which destroy the attempted isolation of a quark.  
Put another way, there is no order parameter for deconfinement.  In a theory without dynamical quarks, one can measure the free energy of an isolated heavy 
test quark, treated as an external source. In such a theory, there is a well defined transition to deconfined matter at a temperature of near $300$~MeV, 
since this free energy is measured in lattice Monte-Carlo simulations and is infinite for $T \le 300$~MeV, and finite for higher temperatures.  
For a theory with dynamical quarks, that is for the real world,  the free energy of an isolated heavy quark is not very well defined at finite temperature and density.  
This is because the free energy of a heavy test quark is given by $1/2$ that a heavy quark-light antiquark meson formed together with corresponding meson composed of a heavy antiquark-light quark formed as one tries to separate a heavy quark and a heavy antiquark. In other words it is not possible to separate an isolated heavy quark and heavy antiquark, and the free energy of a heavy quark source  is finite and is given by the free energy of the meson corresponding to the heavy quark and a light antiquark.

Confinement is however well defined in the large number of quark colors limit.  In this limit, we can compute the effect of color screening on the potential for a heavy quark-antiquark pair.  Such screening is ordinarily  given by the effect of quark-antiquark pairs. This can be understood by knowing the Debye mass, or inverse screening length associated with thermal fluctuations in the system.   At zero temperature the system is confined because there are no states of fractional charge.  If we pull a quark-antiquark pair apart, the contribution of quark-antiquark pairs which might screen the potential are suppressed by an inverse power of the number of colors.  (For a review of the rules for counting factors of $N_{\rm c}$, see the papers of 't Hooft and of Witten 
\cite{tHooft:1973alw,Witten:1979kh}.)  The reason that deconfinement might occur in large $N_{\rm c}$ QCD is entirely due to the gluons.  The gluon contributions to the Debye mass squared are of order $M_{\rm Debye}^2 \sim \alpha_{\rm 't Hooft} T^2$ and when
$M_{\rm Debye}^2 >> \Lambda^2_{\rm QCD}$, Debye 
screening associated with these gluons will cutoff the potential before one can separate a quark-antiquark pair to the distance $1/\Lambda_{\rm QCD}$.

In the large $N_{\rm c}$ limit, it is possible to define the confinement-deconfinement transition, so if $N_{\rm c} =3$ is a fair approximation to the large $N_{\rm c}$ limit, one might qualitatively be able to distinguish between a confined world and one that is not.  This can be done for example by looking at the exponential of the negative of a free energy of an isolated quark, or looking at the entropy of the system to determine whether the number of degrees of freedom are consistent with a confined world, where $N_{\rm dof} \sim 1$ or a deconfined world where they are of order $N_{\rm dof} \sim N_{\rm c}^2$ corresponding to deconfined gluons.   I will argue in the following that indeed at temperatures above $T \ge 300$~MeV, these quantities are consistent with a deconfined interpretation.  But there is an intermediate region for $160 ~{\rm MeV} \le T \le 300$~MeV, where the number of degrees of freedom is of order $N_{\rm c}$, 
consistent with the  number of quark degrees of freedom, but with gluons bound into glueballs, and where the 
free energy of a quark is large corresponding to the confinement scale.  I will also argue that in this region, chiral symmetry is restored, corresponding to massless baryonic or quark degrees of freedom. 

I will show that 3 dimensional string theory quantitatively reproduces the energy density and pressure of quark and gluon systems at finite temperatures on the lattice, and this observation therefore reinforces the existence of this intermediate phase.  Of course, the three dimensional theory cannot describe the lowest mass states, since in string theory these states are tachyonic.  But in QCD, the string description is valid only for high mass states because these states correspond to color charges far separated, so that a color flux tube dynamically forms between them. For the lowest mass states in QCD, the configurations are not stringy, and the dynamics is different.  Here I will treat the contributions of the lowest mass states by taking their masses from experiment, and use the string theory to calculate only the spectrum of higher mass states. For our purposes, string theory provides an approximate dynamical description of QCD valid for excited states. 

The results presented will provide a quantitative realization of some of the speculations by Glozman and Cohen concerning an intermediate phase with 
restored chiral symmetry~\cite{Cohen:2023hbq}.

\subsection{Thermodynamics and the Large $N_{\rm c}$ Limit}

The large $N_{\rm c}$ limit was applied by Thorn to QCD thermodynamics to argue the existence of a Hagedorn temperature~\cite{Thorn:1980iv}.  
The Hagedorn temperature is a limiting temperature for a thermodynamic system~\cite{Hagedorn:1965st}.  
Thorn's argument was very simple.  At low temperature, the number of degrees of freedom is of order one in the number of colors, because confined states are color singlet.  When there is deconfinement, the number of degrees of freedom is of order $N_{\rm c}^2$.  
Therefore\rr[,] at the Hagedorn temperature, the energy density must jump by order $N_{\rm c}^2$, and as $N_{\rm c}\rightarrow \infty$ this will cost an infinite amount of energy.

This can be understood at finite $N_{\rm c}$ by simple arguments.  First we need to understand the interaction strength of mesons and glueballs, the color singlet degrees of freedom.  Consider a meson current which can be thought of as originating from   quark-antiquark pair.  The expectation value of this current is of order $N_{\rm c}$,
\begin{equation}
< J(x) J(0) > \sim N_{\rm c}
\end{equation}
because this correlation function corresponds to computing a quark-antiquark polarization diagram with current insertions which do not involve color.  The overall quark counting corresponds to a loop of  a quark-antiquark pair.
If we insert an intermediate meson state, we find  a typical matrix element is
\begin{equation}
<0\mid J \mid n> \sim \sqrt{N_{\rm c}}
\end{equation}
Now let us estimate the three meson vertex.  The Feynman diagram corresponds to the coupling to the meson current is a closed loop, which is again proportional to $N_{\rm c}$, so that
\begin{equation}
    g_{\rm 3meson} ~J^3 \sim N_{\rm c}
\end{equation}
or $g_{\rm 3meson} \sim 1/\sqrt{N_{\rm c}}$.
Similar arguments will give $g_{\rm 4meson} \sim 1/N_{\rm c}$, and glueball interaction $g_{\rm 3glueball} \sim 1/N_{\rm c}$ and $g_{\rm 4glueball} \sim 1/N_{\rm c}^2$

These arguments show that the screening of the quark potential due to quarks is of vanishing strength in the large $N_{\rm c}$ limit, and that glueball and meson interactions are suppressed.  
Note that if one considers a color current current correlation function, because $\Tr(\tau^a \tau^b) \sim 1$, where $\tau$ is a generator of SU($N\rr[_c]$), 
the polarization diagram which generates Debye screening for the quark potential is of order 1 in powers of $N_{\rm c}$. 
For gluons, the trace of the current correlation function involves matrices in the adjoint representation, and the current current correlation 
function is of  order $N_{\rm c}$. The strength of these contributions is of order $g^2$ where $g$ is the gauge coupling.  
Since $g^2 = g_{\rm 't Hooft}^2/N_{\rm c}$, the quark contribution to Debye screening is suppressed in the large $N_{\rm c}$ limit, whereas the gluon contribution is not.

One way to understand the deconfinement transition is simply that the Debye screening length associated with gluons becomes of order the QCD scale $1/\Lambda_{\rm QCD}$.  
Another way to understand it is in terms of gluon interactions.  The energy density and pressure may be expanded in powers of the quark 
or gluon density times \rr[the] corresponding interaction amplitudes.  To have a gas of quarks would require the entropy be of order $N_{\rm c}$.  
Two meson interactions are of order $1/N_{\rm c}$ so this occurs when the density of quarks is of order $N_{\rm c}$.  
To generate a gluonic contribution to the pressure requires an entropy density of order $N_{\rm c}^2$, or since gluon interactions are of order $1/N_{\rm c}^2$, the glueball density must be of order $N_{\rm c}^2$

This is the heart of the argument that suggests three separate regions of the phase diagram.  The three regions are where thermodynamic quantities are of order $1$, corresponding to weakly interacting mesons and glueballs, a region where thermodynamic quantities are of order $N_{\rm c}$, corresponding to interacting mesons and non-interacting glueballs and a region where thermodynamic quantities are of order $N_{\rm c}^2$ corresponding to interacting glueballs and quarks.  Alternatively, if the mesons or glueballs are strongly interacting, perhaps we should roughly imagine the first region as confined quarks and gluons, the second as confined glueballs\rr[,] with quark degrees of freedom that of de-confined quarks\rr[,] 
and the third region as deconfined quarks and gluons.  This picture is however not quite correct because in the infrared, the intermediate quark-glueball phase may be confined.

\subsection{The Three Dimensional String and Thermodymamics}

In the large $N_{\rm c}$ limit, when the density of mesons and glueballs is small enough so that interactions may be ignored, the thermodynamics 
is sensitive only to the spectrum of particle states.  We simply sum over states with weights corresponding to Bose-Einstein distributions for bosons and Fermi-Dirac distributions for fermions.  The decay widths of such particles may also be ignored since they are suppressed by factors of $1/N_{\rm c}$.  The low lying states in each sector are the kaon and pion Goldstone modes for mesons and the lowest mass glueball states.  Excited states correspond to higher energy states generated when quark or gluon sources are at larger separation inside a meson or a glueball.

The spectrum of these higher mass states can be computed in string theory~\cite{Green_Schwarz_Witten_2012}.  This is string theory for 3 spatial dimensions, corresponding to a physically stretched flux tube of a quark-antiquark pair at the open ends of the string, for mesons, or  a closed loop of gluonic flux corresponding to a glueball.

String theory was originally formulated to describe hadronic physics.  It rapidly became clear that string theory by itself could not fully describe hadronic physics since it had tachyons, or ghosts, in the low mass spectrum.  This is solved in 26 dimensions or when including supersymmetric fermion strings in 10 dimensions.  For our purposes, we take string theory to describe excited states, and treat the lowest mass sates by including their effect directly, and then using the string to compute efects of high mass  states.  As such, our string theory must be in three spatial dimensions.  

For small mass quarks and gluons, the physical parameter which sets the scale for the computation of physical quantities is the string tension, which we will take to be
\begin{equation}
\sqrt{\sigma}  = 440-485~ {\rm MeV}\, .
\end{equation}
The density of states for mesons corresponding to  single flavor and quark-antiquark
\cite{Green_Schwarz_Witten_2012}
\begin{equation}
    \rho_{\text{open}}(M) = \frac{\sqrt{2\pi}}{6 T_{\text{open}}} \biggl(\frac{T_{\text{open}}}{M}\biggr)^{3/2} e^{M/T_{\text{open}}}\,.
    \label{eq:rhoopen}
\end{equation}
The open string formula must be multiplied by a degeneracy factor of the number of quark flavors times spin squared.  
For mesons composed of two light quarks, this factor is 16, for singly strange $\mid S \mid = 1$ mesons, it is 16, and for mesons with a strange-antistrange pair it is 4.
For closed string corresponding to glueballs, the density of states 
\begin{equation}
    \rho_{\text{closed}}(M) = \frac{(2\pi)^3}{27 T_{\text{closed}}} \biggl(\frac{T_{\text{closed}}}{M}\biggr)^4 e^{M/T_{\text{closed}}}\,.
    \label{eq:rhoclosed}
\end{equation}

When computing thermodynamics, one integrates over the masses of various states times a Boltzman factor $e^{-E/T}$.  
This integration does not converge  when the temperature is above the Hagedorn temperature.  
This is because the density of states grows so rapidly that the thermal Boltzman factor, $e^{-M/T}$ cannot make the integral converge.  The Hagedorn temperature is given by
\begin{equation}
    T_{\rm H} = T_{\rm closed} = T_{\rm open} = \sqrt{3\sigma/2\pi} = 304-335~{\rm MeV} \, .
\end{equation}

This Hagedorn or limiting temperature seems too high to be consistent with the accepted lore about the meson spectrum, and the Hagedorn temperature 
in phenomenology is typically taken to be much smaller.  However, these descriptions of the meson spectra did not use the 3 spatial dimensional string  formula for the hadron spectrum, 
with the proper prefactor involving powers of $M/T$. We will see that with the values of the string tension taken from phenomenology, 
one gets a very good description of available data about the density of glueball and meson states, and the thermodynamics of QCD for mesons, 
and for glue in theories without dynamical quarks.  I will show how well the three dimensional string theory describes the spectrum of states a QCD thermodynamics in the following sections~\cite{Fujimoto:2025sxx,Marczenko:2025nhj}.

\subsection{Spectrum of Mesons and Glueballs}
As shown by Meyer, the lattice data on the glueball spectrum and thermodynamics of QCD in the absence of quarks are well described 
by the closed string density of states 
\cite{Meyer:2004gx,Meyer:2009tq}.  In Ref. \cite{Marczenko:2025nhj}, we plotted the formula for the closed string integrated density 
of states and compared it to the lattice Monte Carlo computations  of Ref.~\cite{Athenodorou:2020ani}.  These computations were for a theory with no dynamical quarks.  
It has proven difficult to extract the glueball spectrum in lattice computations in theories with dynamical quarks, since these states are unstable, but we might expect small modification of the integrated spectrum of states at large $N_{\rm c}$.  This is because the widths of such states are small in the large $N_{\rm c}$ limit.
\begin{figure}[t!]
    \includegraphics[width=.5\linewidth]{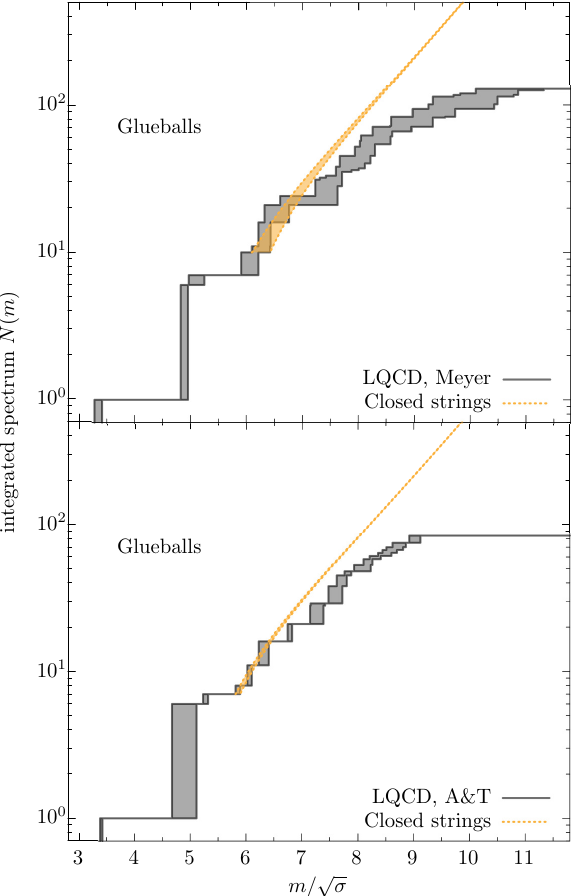}
    \caption{\footnotesize Continuum-extrapolated cumulative mass spectra of glueballs from LQCD simulations (black, solid bands). 
    The spectra are taken from Ref.~\cite{Meyer:2004gx} (top panel) and~\cite{Athenodorou:2020ani} (bottom panel)\rr[,] 
    and are depicted in the units of the string tension $\sqrt\sigma$. The bands represent the uncertainties of the continuum-limit extrapolation. 
    The spectra for closed strings are shown as orange, dashed bands. Their uncertainties come from the uncertainties of the continuum-limit 
    extrapolation of the mass of the lightest resonance in the LQCD spectra (see text).  Figure from Ref. \cite{Marczenko:2025nhj}.}
    \label{fig:spectrum_gb}
\end{figure}

We computed the glueball integrated mass spectrum from the string theory, and separated the lowest mass glueball state and added that in separately.  
The lattice computation specifies a string tension that determines the  Hagedorn temperature $T_{\rm H}/\sqrt{\sigma} = \sqrt{3/2\pi}$.  
The lattice computation no doubt under-counts the glueball states that are above threshold for decay into two gluon states.  
Nevertheless, the three spatial dimensional string provides a zero parameter description of this data, and it describes the lattice data remarkably well, 
as shown in Fig. \ref{fig:spectrum_gb}.

A further test of the string description of QCD in the absence of dynamical quarks is comparison with lattice Monte-Carlo data on thermodynamic quantities.  
To compute the energy density associated with a density of states $\rho(M)$, we can compute for example the energy density as
\begin{equation}
    \epsilon = \int~dM~ \rho(M) \int{{d^3p} \over {(2\pi)^3}} E(p) e^{-E(p)/T}
\end{equation}
where $E(p) = \sqrt{p^2+M^2}$.  We see that the integration over the spectrum of masses does not converge for $T > T_{\rm H}$, 
and that $T_{\rm H}$ is the limiting temperature.  ]If the entropy density $s=dP/dT$
diverges like some power of $1/(T_{\rm H}-T)$, near the Hagedorn temperature, then the pressure is one less power singular.  Because $P = -\epsilon +s$, the energy density will also be more singular than the pressure.  This means that near the Hagedorn temperature, $s \sim \epsilon$, and the contributions to the integration over $M$ is non relativistic.

Note that the pressure is determined from knowing the trace anomaly $\epsilon - 3P$, where $\epsilon$ is the energy density and $P$ is the pressure.  
This is because $dP/dT = s$, where $s$ is the entropy density.  We see that
\begin{equation}
  T {d \over {dT}} {P \over T^4} = {{\epsilon - 3P} \over T^4}
\end{equation}

\begin{figure}[t!]
    \includegraphics[width=.5\linewidth]{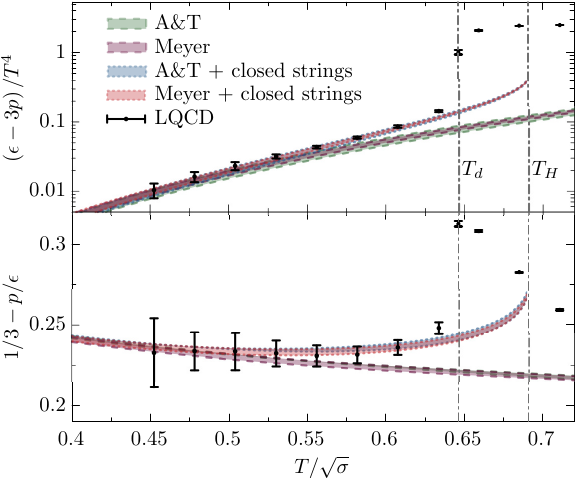}
    \caption{\footnotesize Trace anomaly $(\epsilon - 3p)/T^4$ (top panel) and energy-density-normalized trace anomaly $1/3-p/\epsilon$ (bottom panel). 
    The SU(3) pure gauge LQCD data on pressure and energy density are taken from Ref.~\cite{Borsanyi:2012ve}. 
    The uncertainty bands in both panels are obtained by propagating the reported error on pressure and energy density. 
    The vertical lines $T_{\rm d}/\sqrt{\sigma}=0.646$ and $T_{\rm H}/\sqrt{\sigma}=0.691$ represent the critical deconfinement 
    and Hagedorn temperatures, respectively. Note that the closed strings are shown only up to $T_{\rm H}$ (see text for details). 
    Figure from Ref. \cite{Marczenko:2025nhj}}
    \label{fig:thermo_glue}
\end{figure}
Since different lattice computations use slightly different values of the string tension $\sigma$, it is useful to plot the lattice data for 
 ${{\epsilon - 3P} \over T^4}$, a scale invariant quantity, in terms of the scaled variable $T/\sqrt{\sigma}$.  
 Computations with different values of $\sigma$ should collapse to one plot in this scaling plot.

 The purely gluonic theory for $N_{\rm c}= 3$ has a deconfinement temperature $T_{\rm d}/\sqrt{\sigma} = 0.646$.  
 In the large $N_{\rm c}$ limit, this will rise to the Hagedorn temperature, but it is not far off for $N_{\rm c} = 3$.  Below the deconfinement temperature, 
 the resonance gas description works very well.  Note that there is no divergence of thermodynamic quantities at the Hagedorn temperature, 
 since for purely gluonic theory, the prefactor of $M^{-4}$ \ref{eq:rhoclosed})  is sufficient to guarantee convergence for the entropy density.  
 This is not the case for open strings corresponding to mesons, which has a pre-factor that scales as $M^{-3/2}$ (\ref{eq:rhoopen}).  To see this,
 we can use the non-relativistic approximation for the contribution of the very high mass states near $T_{\rm H}$.  The entropy density 
 for a non-relativistic gas scales as $M^{5/2}$, so that the integration over states goes as $\int ~ dM/M^{3/2}$.  
 For open strings, corresponding to mesons, it diverges as $\int M dM$.  The agreement of the Hagedorn model 
 and the lattice computations is again remarkable for temperatures below the deconfinement transition, as is shown in Fig. \ref{fig:thermo_glue}

\begin{figure}[t!]
    \includegraphics[width=.4\linewidth]{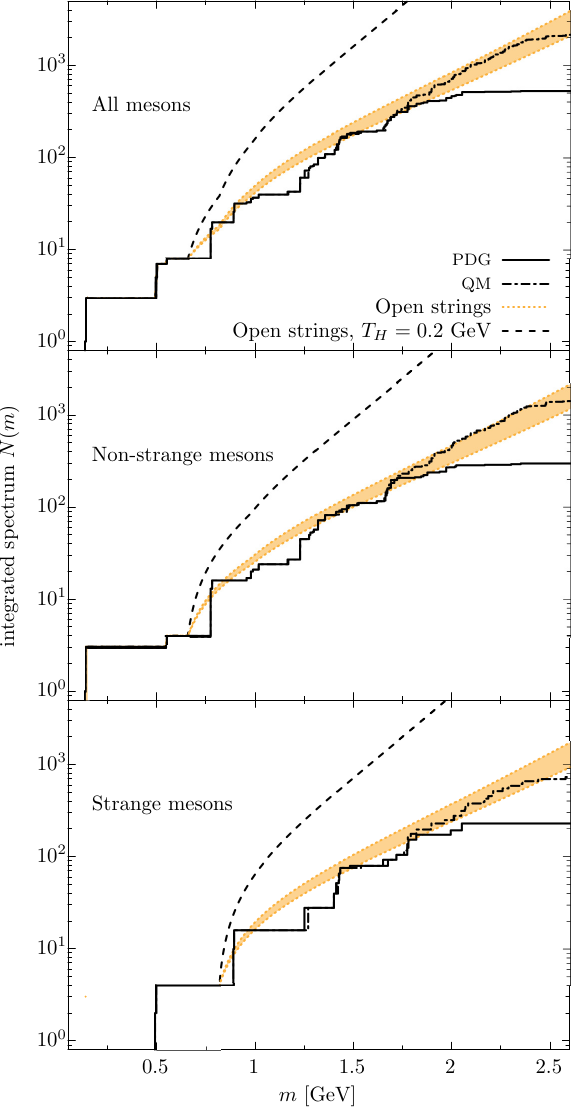}
    \caption{\footnotesize Cumulative mass spectra of all (top), non-strange (middle), and strange (bottom) mesons in the PDG (black, solid lines). Also shown are spectra that include the prediction from the quark model~\cite{Loring:2001ky, Ebert:2009ub} (QM) (black, dash-dotted lines). The yellow bands represent the uncertainty in the Hagedorn limiting temperature in the exponential spectrum (see text for details). We note that $f(500)$ and $\kappa^\star_0(700)$ mesons are not included in the discrete spectra due to their ambiguous nature~\cite{Broniowski:2015oha, Friman:2015zua}. The black, dashed lines show spectra obtained for $T_{\rm H} = 0.2~$GeV. Figure from Ref. \cite{Marczenko:2025nhj}}
    \label{fig:spectra_meson}
\end{figure}

The spectrum of mesons can also be computed.   The region of applicability of the string theory density of states is above that of the 
Goldstone boson states in the sectors of the theory corresponding to the specified strangeness content.
 In Fig. \ref{fig:spectra_meson}, comparison of the computation using the three spatial dimension string theory to the observed spectrum 
 of light mass mesons is shown. Note that in the upper range of masses\rs[,] the experimental measurement\rr[s] presumably 
 miss some states. Figure \ref{fig:spectra_meson} also shows the results of various 
 model computations that should include these missing states~\cite{Loring:2001ky, Ebert:2009ub}.  
 The result\rs[s] of this essentially zero free parameter description is stunning.
 
 \begin{figure}[t!]
    \includegraphics[width=.4\linewidth]{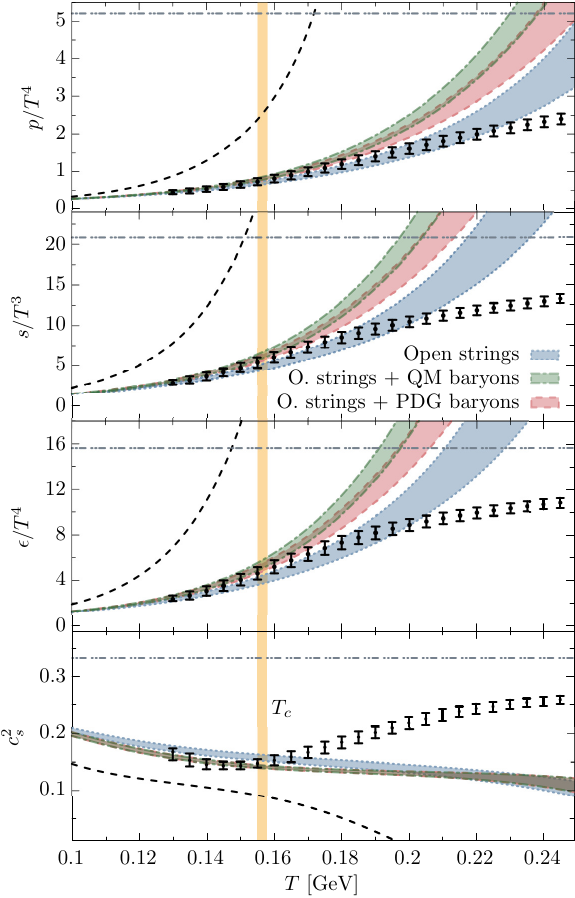}
    \caption{\footnotesize The equation of state at finite temperature and vanishing chemical potential. The LQCD results are taken from Ref.~\cite{HotQCD:2014kol}. 
    Mesons are modeled via the continuous spectrum of open strings, and baryons are taken as discrete states from the 
    Particle Data Group (PDG) or quark model (QM). The blue, red, and green bands indicate the uncertainty in the string tension (see text). 
    The yellow vertical band marks the estimation of the pseudocritical temperature for the chiral crossover transition 
    $T_{\rm \rr[ch]}=156.5\pm1.5~$MeV~\cite{HotQCD:2018pds}. The gray, horizontal, doubly-dotted--dashed lines mark the Stefan-Boltzmann 
    limit considering quarks of 3 flavors and gluons. The black, dashed lines show results for open strings with $T_{\rm H} = 0.2~$GeV. 
    Figure from Ref. \cite{Marczenko:2025nhj}.}
    \label{fig:thermo}
\end{figure}

Finally, we compare our results from the string theory with lattice computations of the energy density, entropy density, 
pressure and the sound velocity, as shown in Fig. \ref{fig:thermo}.  The temperature at which chiral symmetry is restored on the lattice 
is taken to be $T_{chiral} \sim 155$~MeV.  
This is also the temperature for which the meson densities become of order $N_{\rm c}$, as we will see in the following paragraphs.  
The agreement between the string computations and the lattice gauge theory results is again, remarkable. 

\subsection{Chiral Symmetry Restoration}

To compute the chiral condensate from the resonance gas sum, we use
\begin{equation}
<\overline \psi \psi> = <\overline \psi \psi>_0 - {{\partial P} \over {\partial m_Q}}
\end{equation}
The subscript $0$ means vacuum expectation value.
This follows from the definition of the pressure as $P = \beta V \ln( Z/Z_0)$, where $Z$ 
is the thermal expectation value of 
$ e^{-\beta H}$ scaled by $Z_0$ so that the pressure vanishes in vacuum.  
The mass term in the Hamiltonian is $m_Q\overline \psi \psi$.  
To compute the condensate from the resonance gas spectrum, we need to know $\partial M/\partial m_Q$, where $M$ is a meson mass.
For pions, we use the Gell-Mann-Renner-Oakes relationship
\begin{equation}
f_\pi^2 m_\pi^2 = - m_Q <\overline \psi \psi>_0 \, .
\end{equation}
We define $\sigma_\pi = m_Q \partial M /\partial m_Q$ to derive
\begin{equation}
    \sigma_\pi = m_\pi/2 \, .
\end{equation}
For mesons, we will take
\begin{equation}
   \sigma_M = n_q \overline \sigma
\end{equation}
to be proportional to the number of light quarks in a hadron, and to be independent of mass.  
Here $\bar{\sigma}$ is the averaged sigma term.

The pressure $p(M)$ of an ideal meson gas with mass $M$ at zero chemical potential is
\begin{equation}
    p(M) = - T \int \frac{d^3 k}{(2\pi)^3} \ln \left[1 -\exp\left({-\sqrt{k^2 + M^2}/T}\right)\right],
\end{equation}
whose  derivative with respect to $m_q$ can be calculated as
\begin{align}
    \frac{\partial p(M)}{\partial m_q}
    &= -\sigma_M\,\frac{M}{m_q}\frac{1}{2\pi^2}\int_0^\infty dk \frac{k^2 }{\sqrt{k^2+M^2}}
    \frac{1}{e^{\sqrt{k^2 + M^2}/T}-1} \notag\\
    &= -\sigma_M\,\frac{M}{m_q}\frac{TM}{2 \pi ^2}\sum_{n=1}^\infty\frac{1}{n } K_1\left(\frac{n M}{T}\right)\,.
    \label{eq:delP/delmq}
\end{align}

This Nambu-Goldsone (NG) meson contribution leads to~\cite{Gasser:1987ah,Gerber:1988tt}
\begin{equation}
\label{eq:chiralT}
   \frac{\langle\bar{\psi}\psi\rangle_T}{\langle\bar{\psi}\psi\rangle_0}
   \simeq 1-\frac{1}{8}\frac{T^2}{f_\pi^2}\,,
\end{equation}
where we took the chiral limit ($m_q\to 0$) and assumed that there are three pions as the NG mesons.
Since $f_\pi^2\sim \Nc$ in the large-$\Nc$ counting, one may think that Eq.~\eqref{eq:chiralT} implies the critical temperature for vanishing $\langle\bar{\psi}\psi\rangle_T$ is $T_\mathrm{chiral}\sim f_\pi\sim \sqrt{\Nc}$ which diverges for $\Nc\to\infty$.
However, this is not the case in QCD with the Hagedorn spectrum~\cite{Biswas:2022vat}.
The heavier hadrons all contribute to the chiral condensate, which reduces $T_\mathrm{chiral}$.

\begin{figure}
    \includegraphics[width=0.7\linewidth]{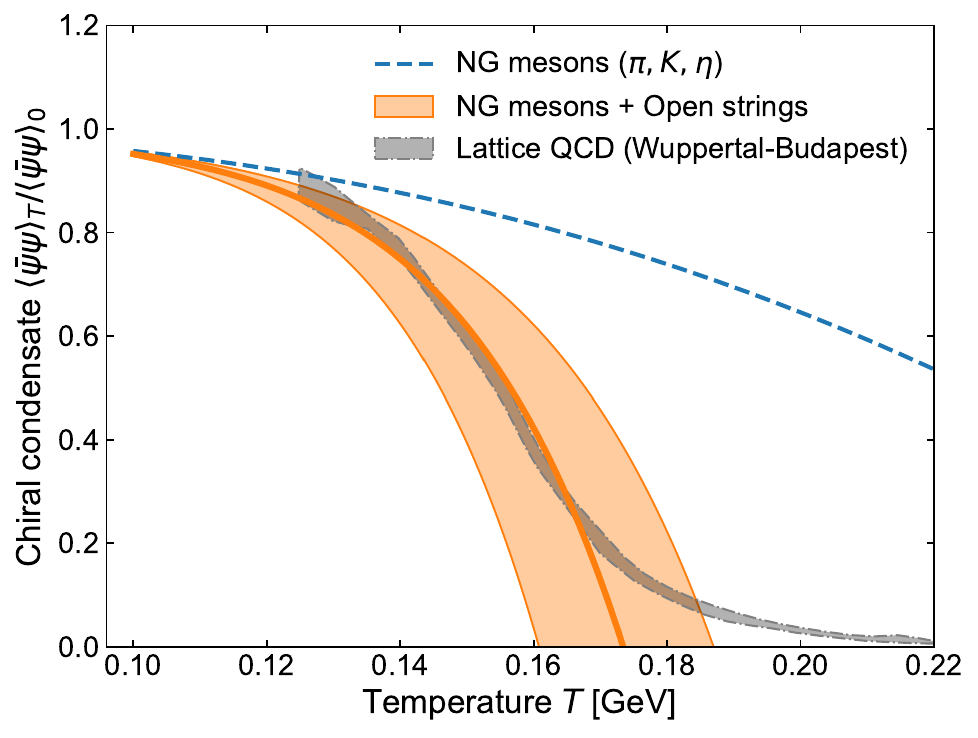}
    \caption{\footnotesize Ratio of the chiral condensate at finite temperature to that at zero temperature.
    The lattice data are taken from Ref.~\cite{Borsanyi:2010bp}.
    The averaged sigma term per quark, $\bar{\sigma} \simeq 30~\mathrm{MeV}$, 
    is obtained from the best fit to the lattice-QCD data and it is varied 
    within $15$ – $60\MeV$ to account for uncertainty.
    The blue curve corresponds to the first term in the square bracket in Eq.~\eqref{eq:chiralT2}, 
    and the orange curve includes the contribution from the second term in the square bracket in Eq.~\eqref{eq:chiralT2}.
    Figure from Ref. \cite{Fujimoto:2025sxx}.}
    \label{fig:chiral_condensate}
\end{figure}

The correction to $\langle \bar{\psi} \psi \rangle_T$ from an open string can be computed
by collecting the contributions from NG mesons and open strings. The thermal correction to the chiral condensate is
\begin{align}
    \frac{\langle\bar{\psi}\psi\rangle_T}{\langle\bar{\psi}\psi\rangle_0} = 1 + \frac{m_q}{m_\pi^2f_\pi^2} \left[
    \sum_{i=\pi,K,\eta} g_i \frac{\partial p_i (M_i)}{\partial m_q}
    +\frac{\partial p_\text{open string}}{\partial m_q} \right]\,,
    \label{eq:chiralT2}
\end{align}
where $p_i$ and $g_i$ denote the pressure and the spin-flavor degeneracy, respectively, of the $i$-th hadron species with mass $M_i$, and we used the Gell-Mann--Renner-Oakes relation to replace $\langle\bar{\psi}\psi\rangle_0$ with $-m_q /(m_\pi^2 f_\pi^2)$ in the above expression.
In Fig.~\ref{fig:chiral_condensate}, we plot the resulting chiral condensate as a function of the temperature.
The sigma terms of the NG mesons are taken to be the same as those in Ref.~\cite{Biswas:2022vat}, and the lattice data are taken from Ref.~\cite{Borsanyi:2010bp}.  

The agreement with such a simple computation is again remarkable.  Note that this occurs when particle densities are generically of order $N_{\rm c}$.

\subsection{Properties of the Intermediate Phase \label{ssec:props}}

\begin{figure}
    \includegraphics[width=0.7\linewidth]{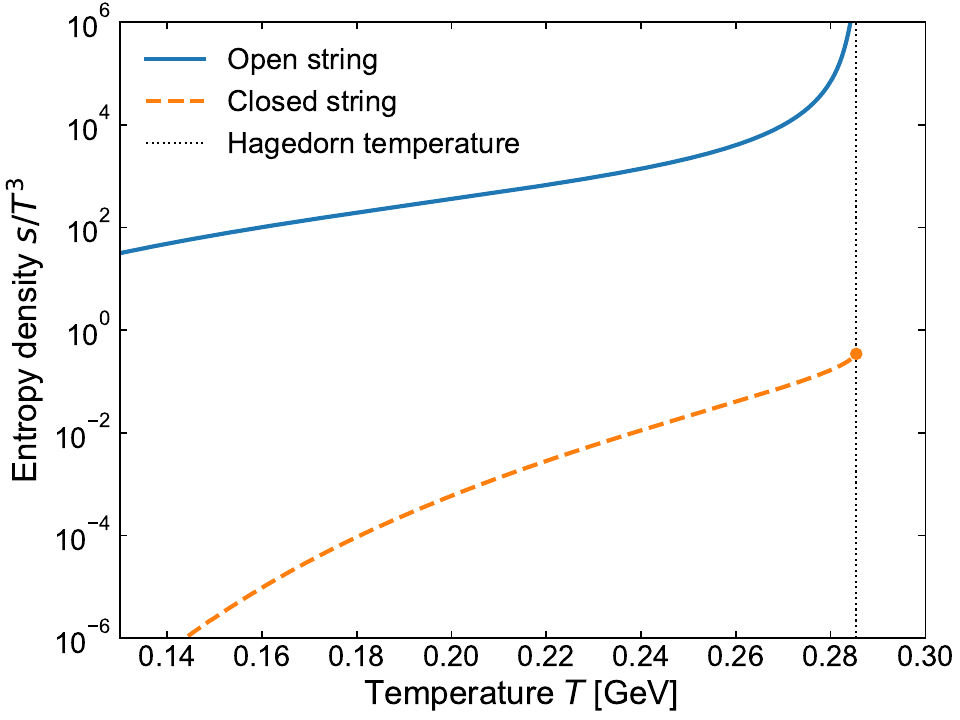}
    \caption{\footnotesize Comparison between the entropy densities from an open-string gas (drawn by the solid curve) and a closed-string gas (drawn by the dashed curve) up to the Hagedorn temperature of $\Th = 285\MeV$. Figure from Ref. \cite{Fujimoto:2025sxx}.}
    \label{fig:openclosed}
\end{figure}

In the discussion above, I argued that there were three phases where the energy density is of order 1, $N_{\rm c}$ and $N_{\rm c}^2$.  In Fig. \ref{fig:openclosed}, 
the contributions to $s/T^3$ from open and closed string are plotted.  The intermediate region is from the chiral restoration temperature, 
$T_{\rm ch} \sim 165$~MeV up to the Hagedorn temperature.  For $T \le T_{\rm ch}$, the system is a confined hadron gas.  Above the Hagedorn temperature, it is a deconfined quark gluon plasma.  From the figure, we see that in the intermediate region, the contribution of glueballs is infinitesimal, so we should think of the system as quarks with only a very small amount of glueballs, since the entropy density scales approximately with $N_{\rm c}$.  Chiral symmetry is restored in this phase.

The disappearance of confinement is usually associated with the Debye screening mass exceeding the QCD scale $\Lambda_{\rm QCD}$.  
In other words, the screening length becomes less than the typical QCD length scale.  The Debye screening mass is 
\begin{equation}
    M^2_\mathrm{Debye} \sim g^2\biggl({1 \over 3} \Nc + {1 \over 6}\Nf\biggr)T^2\,,
    \label{eq:Debye}
\end{equation}
In the 't Hooft limit, $g^2_{\rm 't\,Hooft} = g^2 N_{\rm c}$, so that the second term associated with the quarks vanishes in this limit.  
The first term comes from gluons, and because the abundance of glueballs is small,
this term should probably be ignored in the intermediate phase.  Therefore in the large $N_{\rm c}$ limit, the intermediate phase is confined.  
The quarks must be propagating attached to confining strings, although
the number of degrees of freedom is $N_{\rm c}$ suggesting that the quarks typical kinetic energy is not so strongly affected by confinement.  
For this reason we refer to this intermediate phase as Quark Spaghetti with Glueballs, or SQGP.
\begin{figure}
    \includegraphics[width=0.7\linewidth]{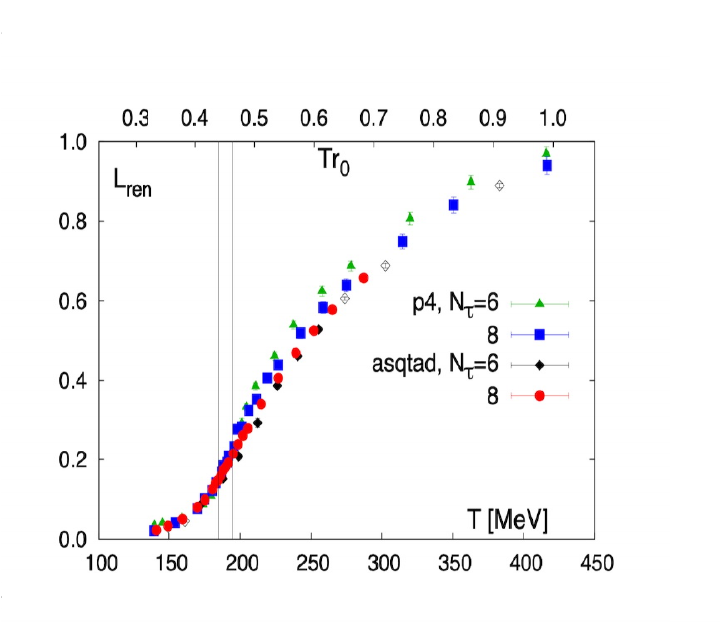}
    \caption{\footnotesize The expectation value of the Wilson line which measure the exponential of the free energy of an isolated quark, $e^{-\beta F_Q}$ 
    as a function of temperature, from Karsch et. al.\cite{HotQCD:2014kol}
    }
    \label{fig:confinement}
\end{figure}

For $N_{\rm c} = 3$ and $N_{\rm F} = 2-3$, corresponding to QCD, it is not so clear where deconfinement occurs.  In
Fig.~\ref{fig:confinement}, I show the expectation value of the Polyakov loop as a function of temperature for the QCD lattice simulation of \cite{HotQCD:2014kol}.  
This expectation value measures the free energy of an isolated quark and reads
\begin{equation}
    <L> = e^{-\beta F_Q} \, .
\end{equation}
\begin{figure}
    \includegraphics[width=0.7\linewidth]{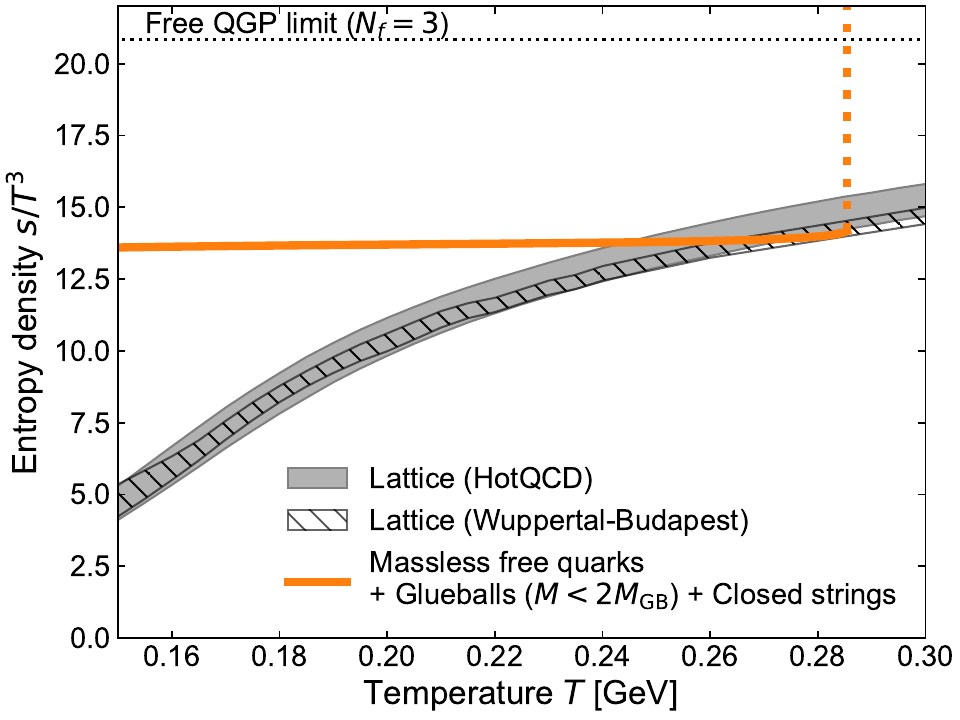}
    \caption{\footnotesize Comparison of the entropy density between the lattice-QCD data and the gas of massless 
    free quarks and light glueballs as well as heavy glueballs that obey the closed-string spectrum. Figure from Ref. \cite{Fujimoto:2025sxx}.}
    \label{fig:closed}
\end{figure}
When $<L>$ is close to one, we expect the system to be more deconfined and for $<L> \sim 0$, it is more confined.  For $T \sim 165 $~MeV, 
$<L> \sim 0.1$ and for $T \sim 300$~MeV, $<L> \sim 0.8$.  So the transition between confinement and deconfinement of quarks appears in the SQGB. 

Note that the appearance of glueballs in the SQGP is strongly suppressed, but of course in the real wold there will be some 
mixing between glueball states and meson states, suppressed in the large $N_{\rm c}$ limit.  
The Hagedorn temperature is a reminder of this dis-association temperature and a measure of when the system may be thought of as a quark gluon plasma.

Another measure of deconfinement is the entropy, $s/T^3$.  In Fig. \ref{fig:closed}, the entropy  computed from lattice gauge theory is compared to that of a massless gas of quarks.  The gluon contribution is omitted.  This free massless quark  entropy agrees with the lattice data very well in the temperature
range of  $240-300$~MeV.  This suggests that
 for bulk thermodynamic quantities  in this upper temperature range a free gas equation of state is not such a bad  approximation. 
 In the lower temperature range, presumably meson interactions are becoming of increasing importance 
until the free gas limit is reached.  This observation suggests that deconfinement of quarks may appear at the temperature appropriate for the SQGB.  
This of course does not mean the system is deconfined, but simply that the kinetic energy terms in the QCD Hamiltonian, 
which measure typical quark kinetic energies, are close to those of free quark values, and that interactions may be not so important here.  
Nevertheless, there will remain quantities sensitive to the infrared\rr[,] which may not be so easily explained.

\section{Lecture 2:  The Phase Diagram at Finite Temperature and Baryon Number Density}

\subsection{Introduction}

Neutron stars provide a good determination of the equation of state of nuclear matter.  For a recent review of the status 
of such  determination and a comprehensive set of references, see Ref. \cite{Fujimoto:2024cyv}, and in particular 
the seminal theoretical analysis of Refs. \cite{Annala:2017llu,Baym:2017whm}.  
These works infer that the equation of state is stiff at densities a few times that of nuclear matter with the sound velocity squared of the order of and possibly exceeding 
${1 \over 3} c^2$ \cite{Bedaque:2014sqa,Kojo:2014rca,Tews:2018kmu}. This is a truly remarkable observation since at nuclear matter density, 
the degrees of freedom are non-relativistic nucleons, with binding energies and kinetic energies small compared to the nucleon mass. 
Apparently if one squeezes nuclear matter increasing the density by a few times, the system rapidly becomes relativistic.  
This observation hints that the transition may be from nucleons to  quarks.  However, this is unlikely to be a first or second order phase transition 
since at such transition the sound velocity drops to zero and the equation of state becomes soft.

The subject of this lecture will be to show how a rapid transition to a stiff equation of state might occur if matter is Quarkyonic.  
Such Quarkyonic matter was proposed to explain a remarkable fact observed in the large $N_{\rm c}$ limit of QCD \cite{McLerran:2007qj}.  
At large $N_{\rm c}$, de-confinement occurs at a quark chemical potential (or Fermi energy) that is parametrically large compared to the QCD scale, 
$\mu_{\rm quark} \sim \sqrt{N_{\rm c}} \Lambda_{\rm QCD}$.  This follows because the Debye mass at finite density is $M_{\rm D}^2 \sim \mu_Q^2/N_{\rm c}$ 
so that this becomes of the order of the QCD scale when $\mu_Q = \sqrt{N_{\rm c}} \Lambda_{\rm QCD} >> \Lambda_{\rm QCD}$. 
The deconfnement transition happens at momentum scales parametrically large compared to the QCD scale.  
On the other hand, we might  naively expect that the effects of interactions will be small when $\mu_{\rm quark} >> \Lambda_{\rm QCD}$.  
The resolution of this paradox is that at the Fermi surface, there are strong interactions since Fermi surface interactions are not cutoff by Debye screening. 
The Fermi surface may be thought of as confined baryons with confined mesonic excitations.  On the other hand, deep inside the Fermi surface, interactions 
are weak and the degrees of freedom may be thought of as those of quarks.  Since the bulk properties of the system should be largely determined by the bulk 
of particles which are in the Fermi sea, the system becomes relativistic at a scale $\mu_{\rm quark} \sim \Lambda_{\rm QCD}$, and this corresponds to a density scale 
not so large compared to that of nuclear matter.  There may or may not be a true  phase transition to Quarkyonic matter, but the transition to this matter 
is distinct from the de-confinement transition.  I shall later show that such a  transition can occur at lowish density, and that it allows for a rapid stiffening 
in the equation of state of high density matter.  

In the Quarkyonic region the baryon number density is controlled by the quarks, not the thin shell of baryons near the Fermi surface, $n_B \sim \mu_Q^3$.  
But the energy density is $\epsilon \sim N_{\rm c} \mu_Q^4$.  This was like the case for the SQGB where the energy density is $\epsilon \sim N_{\rm c} T^4$.  
The range of the Quarkyonic region is $\sqrt{N_{\rm c}} \Lambda_{\rm QCD} \ge \mu_Q \ge \Lambda_{\rm QCD}$.

\subsection{Scale Invariance and High Density Matter}

The mass and radius of a neutron star are determined by equating the outward pressure against the inward force of gravity, using the 
Tolman-Oppenheimer-Volkov equation of hydrostatic equilibrium.  The equation of state may be characterized by the sound velocity
\begin{equation}
   v_s^2 = {{dP} \over {d\epsilon}}
\end{equation}
where $P$ is the pressure and $\epsilon$ is the energy density.  
Also the trace of the stress energy tensor
\begin{equation}
  T^{\mu}_\mu = {1 \over 3} \epsilon - P
\end{equation}
which can be characterized by 
\begin{equation}
 \Delta = {1 \over 3} -{P \over \epsilon}
\end{equation}
is a useful measure of the equation of state.
These two dimensionless quantities, $\Delta $ and $v_s^2$ characterize the hardness or softness of the equation of state.  $\Delta \sim 1/3$ and $v_s^2 \sim 0$ are characteristic of soft non-relativistic matter, while $\Delta \sim 0$ and
$v_s^2 \sim {1 \over 3}$ are for hard, relativistic matter.

The relationship to scale invariance is easily seen since for a scale invariant theory at fixed baryon number $N$ with volume $V$, the number density is $n = N/V$, and the energy density is $\epsilon = \kappa n^{4/3}$, where $\kappa$ is a constant.  The pressure is
\begin{equation}
   P = - {{dE} \over {dV}} = -{d \over {dV}} {\kappa N^{4/3} \over V^{1/3}} = {1 \over 3} \epsilon \, ,
\end{equation}
so that
\begin{equation}
 v_s^2 = {1 \over 3}
\end{equation}
and
\begin{equation}
  \Delta = 0 \, .
\end{equation}

In QCD, there is a scale associated with the distance scale of confinement, $1/\Lambda_{\rm QCD}$.  With the beta function of QCD 
defined by $\beta(g) = dg/d\ln(\Lambda_{\rm QCD})$, where $g$ is the QCD interaction strength, deviations from scale invariance are quantified by the trace anomaly
\begin{equation}
T^\mu_\mu = -{{\beta(g)} \over g} (E^2-B^2) + m_q(1+\gamma_q) \overline \psi \psi \, .
\end{equation}
In this equation, $E$ and $B$ are the color electric and color magnetic fields, $m_q$ are quark masses, $\gamma_q$ is an anomalous dimension of Fermion operators, and $\psi$ is a Fermion field.  The $\beta$ function is negative for QCD, and usually the effect of quark masses may be ignored, except for pion physics. For single particle states
\begin{equation}
<p\mid T^\mu_\mu \mid p> \sim p^2 = m^2 \ge 0 \, .
\end{equation}
Ignoring the last term, this equation means that $E^2 > B^2$ inside massive hadrons, which is consistent 
with our intuition for bound states made of massive quarks, where electric fields should be larger than magnetic fields.

For gasses composed of hadrons we might therefore expect that $<T^\mu_\mu >$ is positive, and approaches zero from above  as temperatures and densities increase.  This might be violated if there were condensates or systems dominated by pions where the quark mass term might be important.

\begin{figure}
\begin{center}
\includegraphics[width=0.55\textwidth]{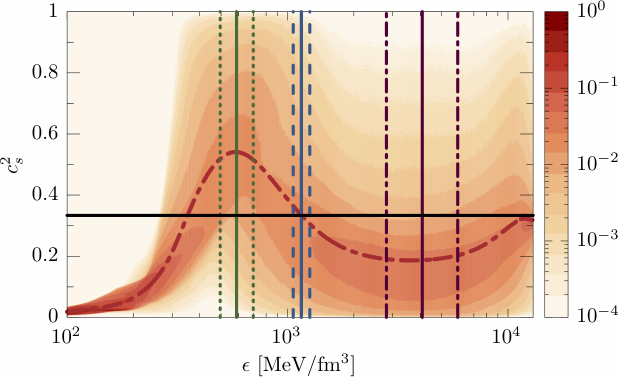}
\end{center}

\caption{\footnotesize The sound velocity as a function of energy density from Ref. \cite{Marczenko:2022jhl}  The green lines indicate where a maximum of the sound velocity is found and the blue lines are typical maximum central densities of neutron stars.  The inference above this maximum density comes from requiring that fits to data smoothly connect to QCD computations at the highest densities.  Figure from Ref. \cite{Marczenko:2022jhl}. }
\label{fig:pd}
\end{figure}

\begin{figure}
\begin{center}
\includegraphics[width=0.55\textwidth]{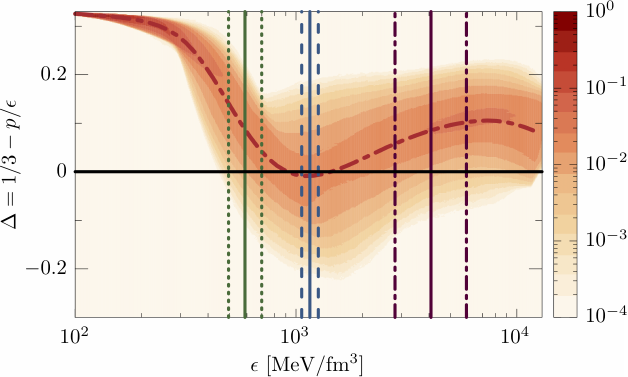}
\end{center}

\caption{\footnotesize The scaled trace anomaly $\Delta$ as a function of energy density from Ref. \cite{Marczenko:2022jhl}} 
\label{fig:pd1}
\end{figure}

One of the remarkable achievements of neutron star studies is that one can extract to a fair approximation the equation of state 
of nuclear matter from neutron star properties \cite{Fujimoto:2024cyv,Annala:2017llu}.  In Fig.~\ref{fig:pd}, the sound velocity extracted from 
Bayesian analysis of  data is shown from Ref.~\cite{Fujimoto:2022ohj,Marczenko:2022jhl},
and it appears that the the equation of state is close to the conformal limit at high density, as shown in Fig.~\ref{fig:pd1}.  
This is a surprise since the energy density is not  so high  compared to the QCD scale. Generally, the sound velocity squared exceeds $1/3$ 
above some density and there is some evidence for a maximum at a density a few times that of nuclear matter.  
The scaled trace of the stress energy tensor approaches zero from above, characteristic of a gas of massive particles

There is a relation between the sound velocity and the trace anomaly since
\begin{equation}
  v_s^2  = -\epsilon {{d} \over {d\epsilon}} \Delta - \Delta + {1 \over 3} \, .
\end{equation}
If the scaled trace  approaches zero monotonically, this equation means that a large sound velocity accelerates this approach, and a large sound velocity is signal for a precocious approach to the conformal limit.

\subsection{Properties of Quarkyonic Matter}

The observation that the sound velocity changes rapidly to a relativistic value as the baryon number density changes by of order one, has strong consequences.  
The sound velocity can be expressed in terms of the baryon density and baryon chemical potential as
\begin{equation}
{{n_B} \over {\mu_B dn_B/d\mu_B}} = v_s^2 \, .
\end{equation}
\rr[This] means that
\begin{equation}
   {{\delta \mu_B} \over \mu_B} \sim v_s^2 {{\delta n_B} \over n_B} \, .
\end{equation}
If the sound velocity is of order one, an order one change in the baryon density generates a change in the baryon chemical potential which is 
of the order the baryon chemical potential.  Initially, the baryon chemical potential for a non-relativistic system is of order the nucleon mass, 
with a small correction due to nucleon binding  and kinetic energies,
\begin{equation}
   \mu_B = M + \mu_B^{\rm kinetic} \, ,
\end{equation}
but after a change of order one, the kinetic energies are relativistic.  If the distribution of particles were uniform in momentum space, we would expect, $k_F \sim M$, 
and the density $n_B \sim k_F^3 \sim M^3$.  If we count factors of the number of colors, the number density would naively be required to jump by of order $N_{\rm c}^3$.  
The jump is only of order $n_B$, and this means that the baryons must appear in some geometrically restricted region of momentum phase space.

\begin{figure}
\begin{center}
\includegraphics[width=0.55\textwidth]{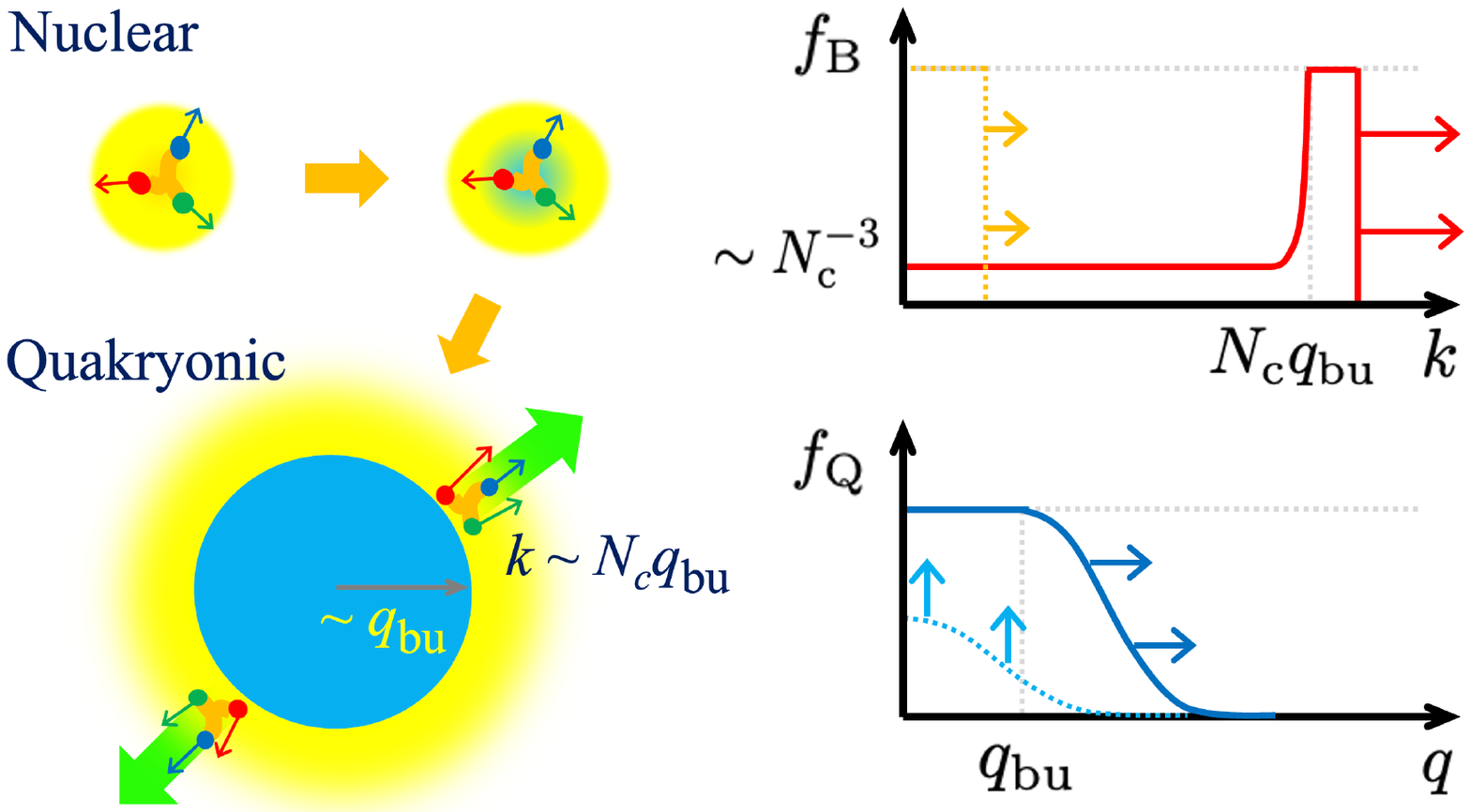}
\end{center}

\caption{\footnotesize The shell structure of Quarkyonic matter from Ref. \cite{Fujimoto:2023mzy}} 
\label{fig:pd4}
\end{figure}

This is solved in Quarkyonic matter.  Nucleons form a shell surrounding  a filled Fermi sea of quarks, as shown in Fig. \ref{fig:pd4}, 
as computed explicitly in Ref. \cite{Fujimoto:2023mzy}.  The nucleonic shell has a thickness that can become thinner as the baryon density 
increases, so that the density of baryons associated with the momentum space  shell does not rapidly grow, in spite of the fact that the typical 
nucleon momentum can grow.  The typical Fermi momentum of the quarks may be as high as $\Lambda_{\rm QCD}$, and the baryon density 
in the quarks is of order $\Lambda_{\rm QCD}^3 $ which is not so high, in spite of the fact that the nucleon momentum scale 
is $N_{\rm c} \Lambda_{\rm QCD} \sim M_N$ .  Making a shell of relativistic nucleons surrounding a filled Fermi sea of quarks can 
therefore be accomplished at a density $n_B \sim \Lambda_{\rm QCD}^3 << M_N^3$
 
 I shall later construct a theory which explicitly has this property, and moreover is dual in its description of nucleons and quarks.  
 Duality is a most interesting fundamental property of QCD meaning that one should be able to think about the degrees of freedom 
 of quarks and nucleons simultaneously because nucleons are composed of quarks, and thinking in terms of quarks or nucleons simply 
 involve an arbitrary choice of basis states \cite{Ma:2019ery,Ma:2021zev}.
 
 It is useful to consider explicitly the relationship between quark and nucleon densities and to make explicit the $N_{\rm c}$ dependence.  
 The relationship between constituent quark masses and nucleon masses is
 \begin{equation}
 m_q = m_N/N_{\rm c}
 \end{equation}
 and between chemical potentials
 \begin{equation}
 \mu_q = \mu_N/N_{\rm c}
 \end{equation}
For Fermi momenta,
\begin{equation}
k_q^2 = \mu_q^2 - m_q^2 = k_N^2/N_{\rm c} \, .
\end{equation}
For an ideal gas of 2 flavors of quarks
\begin{equation}
 n_B^N = {2 \over {3\pi^2}}k_N^3
\end{equation}
and 
\begin{equation}
n_B^q = {2 \over {3\pi^2}}k_q^3 \, .
\end{equation}
The baryon number can be kept from growing too rapidly by adjusting the thickness of the shell.  
The baryon number density of the quarks is naturally $n_B \sim \Lambda_{\rm QCD}^3$ \cite{McLerran:2018hbz}.  
Note that  for a gas of nucleons with Fermi momentum $\Lambda_{\rm QCD}$, the energy density is also of this order, 
so there need not be parametric changes in densities when one makes a transition between a gas of nucleons and Quarkyonic matter.

If we think about a filled Fermi sea of quarks in terms of nucleons, then the corresponding nucleon Fermi momentum for this filled  sea of quarks is $k_N \sim N_{\rm c} k_q$.  
On the other hand, the phase space density of quarks is $f_q \sim  1$, for a fully occupied Fermi sea, but in order that one gets the same baryon number $n_B \sim f_q k_q^3 = f_B k_N^3$,
the occupation number density of nucleons must be $f_N \sim 1/N_{\rm c}^3$.  This is shown in Fig  \ref{fig:pd4}.

Energy densities for the nucleons are $\epsilon_B \sim k_F n_B \sim N_{\rm c} \Lambda^3_{\rm QCD}$ and 
$\epsilon_Q \sim \Lambda_{\rm QCD} n_q \sim \Lambda_{\rm QCD} N_{\rm c} n^q_B$ (since the baryon number per 
quark is of order $1/N_{\rm c}$), so that a change in the energy density  from a nucleon gas to Quarkyonic matter can also be smooth.

The pressure is different however.  For an ordinary gas of nucleons at density $n_B$ the pressure is of order $p \sim k_F^5/M_N \sim \epsilon_B/N_{\rm c}^2$ 
whereas the pressure of a quark gas at this density is of order $P \sim \epsilon_B$.  This means that the equation of state becomes much stiffer when one 
makes a transition between a non-relativistic system of nucleons and a Quarkyonic system, even though the energy density and baryon number density change by of order one.

This is precisely what is needed to describe neutron stars.

\subsection{An Explicit, Exactly Solvable Model for Quarkyonic Matter}

Here I construct the solvable Idylliq model of Quarkyonic matter~\cite{Fujimoto:2023mzy}.  Idylliq is an acronym for ideal Quarkyonic matter.  
As such, I consider a noninteracting gas of nucleons.  I assume that these nucleons are composed of quarks, and one can compute the phase density of quarks, 
$f_q$\rr[,] in terms of the phase space density of nucleons $f_N$.  I will require that the phase space densities satisfy $1 \ge f_q, f_N \ge 0$.  

This is a very simple theory, but we will see that it is a non-trivial theory with two different phases. 
The low density phase is an ideal degenerate Fermi gas of nucleons. At some density of the order of the scale set by the typical momentum 
of a quark inside a nucleon, there is a transition to a high density phase.  This high density phase has a hard equation of state.  
For a particular choice of probability density, this theory is analytically solvable.  The theory has an explicit duality between quarks and nucleons.

Explicitly, the duality relationship between quarks and nucleons is
\begin{equation}
 f_q(\vec{k}) = \int~ d^3p~ K(\vec{k} -\vec{p}/N_{\rm c}) f_N(\vec{p}) \, ,
\end{equation}
where $K $ is a probability distribution for a quark inside the nucleon
\begin{equation}
 \int~ d^3k~ K(\vec{k}) = 1 \, .
\end{equation}
\begin{figure}
\begin{center}
\includegraphics[width=0.75\textwidth]{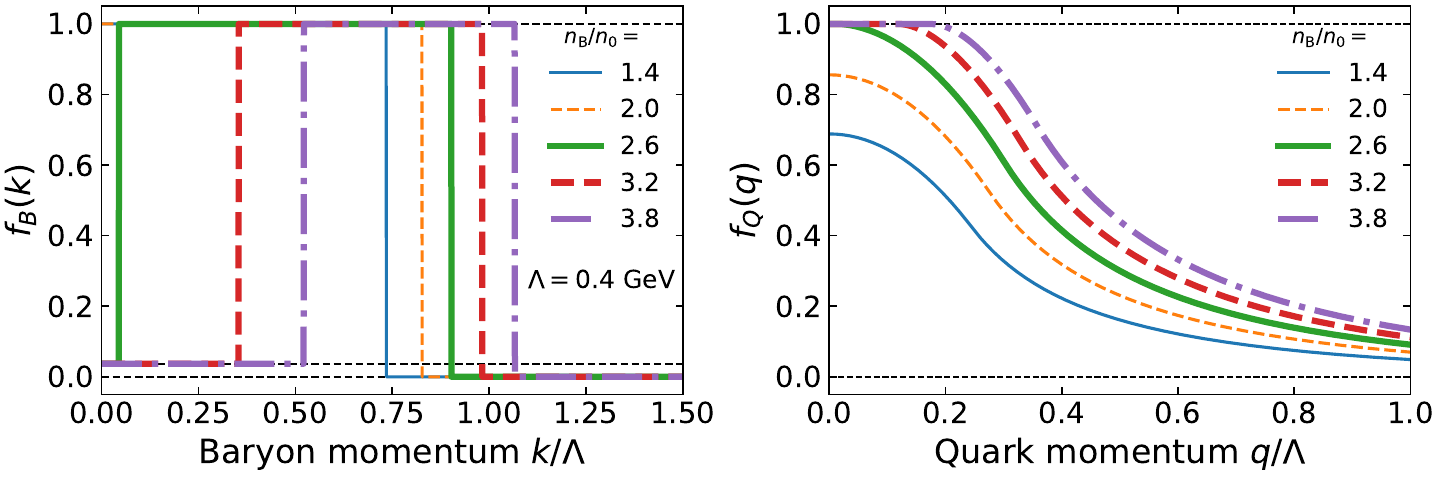}
\end{center}
\vspace{-0.5cm}
\caption{\footnotesize From Ref. \cite{Fujimoto:2023mzy}, the momentum distribution of nucleons (left) and quarks (right) as one increases the baryon number density} 
\label{fig:pd6}
\end{figure}

At low baryon number densities, the constraint that the occupation number density of quarks is small\rr[,] is easy to satisfy.  
This is because the nucleons are localized in low momentum states, but the distribution of quarks is spread out.  
This is shown in Fig. \ref{fig:pd6}.  As the baryon number density increases, more and more nucleon states pile up  for the quark density at zero momentum.  
At some density, the quark occupation phase space density becomes one and beyond this density, the free nucleon 
Fermi gas solution is no longer valid, as shown in Fig. \ref{fig:pd6}.  A formula for this saturation criteria was first proposed by Kojo and is~\cite{Kojo:2021ugu},
\begin{equation}
 1 = \int ~ d^3p ~  K(p/N_{\rm c}) f_N(p)  \label{density} \, .
\end{equation}

The form of solution we will find at higher densities is not so hard to guess.  If nucleons are concentrated on a Fermi surface, and the Fermi momentum increases, this will not so much affect the density of quarks at low momentum.  The quarks on the other hand will begin to fully occupy their phase space, $f_q \sim 1$.  This is the form of the solution in Fig. \ref{fig:pd6}.

To proceed further, we take an explicit form for $K(p)$, that allows for an exact analytic solution,
\begin{equation}
K(\vec{k}) = {1 \over {4\pi \Lambda^2}}{{e^{-\mid \vec{k} \mid}} \over {\mid \vec{k} \mid}} \, .
\end{equation}
This is the Green\rs[s] function for
\begin{equation}
\{ -\nabla^2_k + {1 \over \Lambda^2} \} K(k) = {1 \over \Lambda^2} \delta^{(3)}(\vec{k}) \, .
\end{equation}
Applying this differential operator to the expression for the quark phase space density gives
\begin{equation}
\{ -\nabla^2_k + {1 \over \Lambda^2} \} f_q(k) = {{N_{\rm c}^3} \over \Lambda^2} f_N(N_{\rm c}k) \, .
\end{equation}
This allows us to  determine explicitly the quark distribution in terms of a nucleon distribution. For a distribution of nucleons corresponding
\begin{equation}
f_N(p) = \theta(p_1-p)\theta(p-p_2) + {1 \over {N_{\rm c}^3}}\theta(p_2-p) 
\end{equation}
Here, the momentum $p_1$ is the maximum allowed momentum for all the nucleons, and $p_2$ is the upper momentum for an under-occupied distribuion of nucleons, $0 \le p \le p_2$ with height $1/N_c^3$.  The nucleons between $p_2$ and $p_1$ are fully occupied and may be thought of as a shell of nucleons. 
The  corresponding quark distribution function is much like a Fermi-Dirac distribution, of height one at low momentum, and an exponentially falling tail at the 
Fermi surface, corresponding to a homogeneous solution of the above equation.  The quark Fermi surface is at momentum $k_q = p_2/N_{\rm c}$.

The energy density of a free gas of nucleons can be re-expressed explicitly in terms of a linear theory of quarks by using the relationship between quark and nucleon densities.  
The energy density is
\begin{equation}
\epsilon = \int~ d^3p~ \sqrt{p^2 + M_N^2} f_N(p) = N_{\rm c} \int~ d^3k~E_q(k) f_q(k) \, ,
\end{equation}
where
\begin{equation}
 E_q(k) = \sqrt{k^2+m_q^2} - {{m_q^2 \Lambda^2} \over {(k^2+m_q^2)^{3/2}}} \, .
\end{equation}

In Ref. \cite{Fujimoto:2023mzy}, it is argued that the solution outlined above for the distributions of nucleons and quarks  
is the minimum energy solution at fixed baryon number.  The density at which this solution turns on is when Eq. (\ref{density}) is satisfied,
\begin{equation}
  k_F \sim {\Lambda \over N_{\rm c}^{1/2}} \, .
\end{equation}
The extra factor of $\sqrt{N_{\rm c}}$ arises from the assumed $1/\mid k \mid$ singularity of the probability distribution and might not be present 
in more generic less singular models.  However, this has the feature that the transition from ordinary nuclear matter to Quarkyonic matter happens 
at a very low density compared to the natural QCD scale, and might explain why the transition to Quarkyonic matter occurs at such a low density, very close to that of nuclear matter~\cite{Koch:2024qnz,McLerran:2024rvk}.

\begin{figure}
\begin{center}
\includegraphics[width=0.55\textwidth]{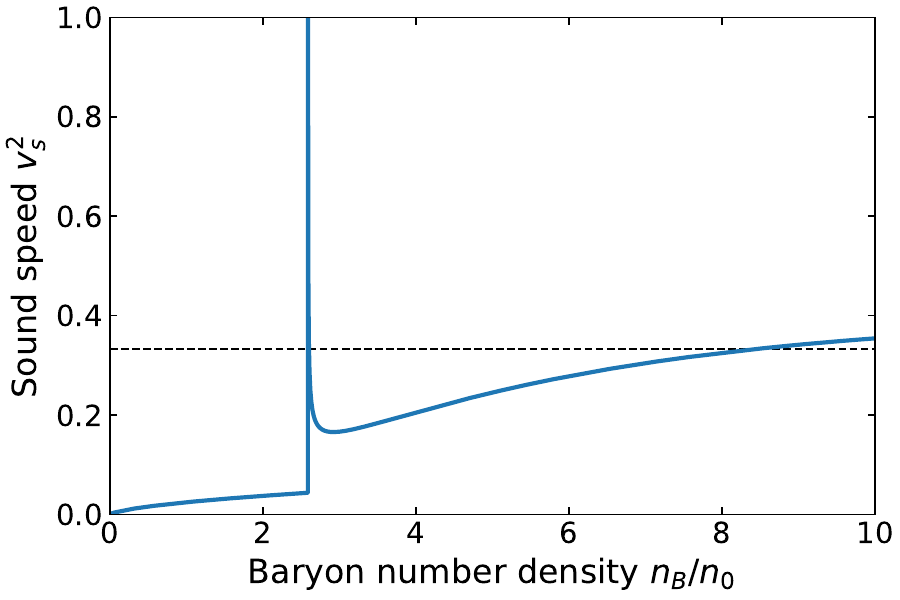}
\end{center}

\caption{\footnotesize From Ref. \cite{Fujimoto:2023mzy}, the sound velocity as a function of density.} 
\label{fig:pd7}
\end{figure}

The transition to Quarkyonic matter in the simple Idylliq model is too rapid, and an explicit computation of the sound velocity shows 
that it has a singularity at the transition \cite{Fujimoto:2023mzy}.  This must be an artifact of ignoring the nucleon interactions, 
which leads to sharp surfaces where the nucleon number is discontinuous.  The region where the discontinuity in the sound speed 
occurs is very narrow in terms of the baryon number density, and becomes zero in the large $N_{\rm c}$ limit, and so should be easy 
to smear out.  A proper theory must remove this singularity, and this is one of defects of the Idylliq model. In Fig. \ref{fig:pd7}, 
a computation of the sound velocity is shown~\cite{Fujimoto:2023mzy}.

\subsection{The Phase Diagram as a Function of Baryon Number Chemical Potential and of Temperature}
\begin{figure}
    \includegraphics[width=0.7\textwidth]{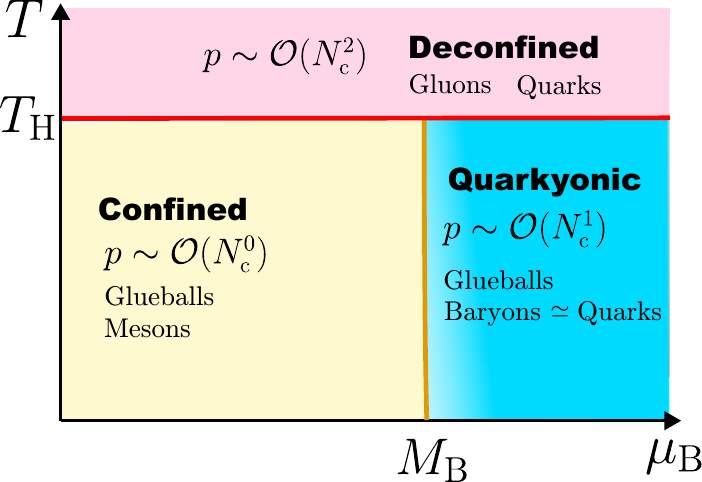}
    \caption{\footnotesize Quarkyonic phase diagram in the strict large-$\Nc$ limit. Figure is from Ref. \cite{Fujimoto:2025sxx}.}
    \label{fig:diagram_largeN}
\end{figure}
At finite temperature and density, we found that in the large but finite $N_{\rm c}$ for fixed number of flavors $N_F$, that there were three phases, the hadron gas, the SQGB, and the QGP.  First consider the limit where $N_{\rm c} \rightarrow \infty$.  In this limit, the temperature for the SQGB phase approaches the Hagedorn temperature. A finite density of baryons does not change this, since quark loops are suppressed  by $1/N_{\rm c}$ until the chemical potential becomes larger than the nucleon mass.  Then one can have a finite density of baryons  The  phase diagram for QCD in this strict large $N_{\rm c}$ limit is shown in Fig.\ref{fig:diagram_largeN}.

\begin{figure}
    \includegraphics[width=0.8\textwidth]{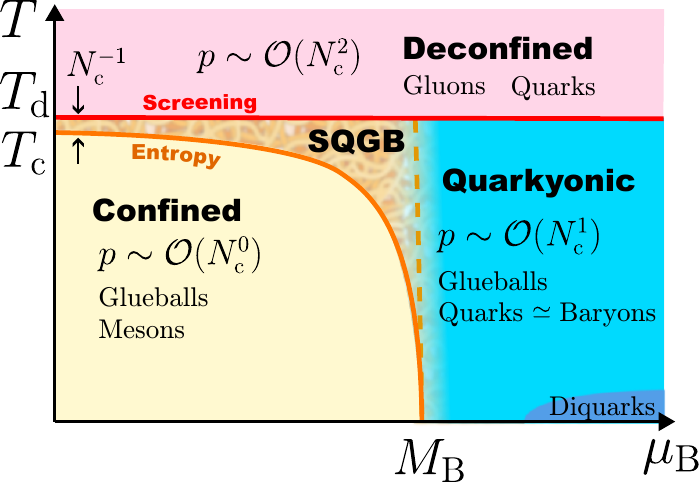}
    \caption{\footnotesize New phase diagram with stretched Quarkyonic Matter toward the temperature axis.  Along the temperature axis, 
    a new window of the SQGB regime can open with a suppressed width $\sim \mathcal{O}(\Nc^{-1})$. Figure is from Ref. \cite{Fujimoto:2025sxx}.}
    \label{fig:diagram_finiteN}
\end{figure}

For large but finite $N_{\rm c}$, the Hagedorn and the SQGB or chiral transition temperature begin to split apart.
There emerges a region  where the energy density scales as $N_{\rm c}$ which joins the $\mu_B = 0$ world with that of finite density.  
It is not clear exactly how this matches onto Quarkyonic matter.  We expect that for the Quarkyonic world, chiral symmetry breaking 
will occur on the Fermi surface, which will disappear as the temperature is raised \cite{Hidaka:2008yy,Kojo:2009ha}.  
The condensation at the Fermi surface is because quark states are occupied in the Fermi sea, and the only available 
unoccupied states for qaurks inside a meson are at the Fermi surface. There is presumably some weak boundary 
between the Quarkyonic world and the SQGB.  A hypothetical phase diagram for this very large but finite $N_{\rm c}$ world is shown in Fig. \ref{fig:diagram_finiteN}.

For realistic $N_{\rm c}$ and $N_F$, we estimated that the temperature at which the chiral temperature is known from the lattice to be of order 
$T_{chiral}\sim 160$~MeV.  The Hagedorn temperature is about $T_{\rm H} \sim 300$~MeV.  We expect the deconfined Quark Gluon Plasma 
occurs at a temperature a little below this.  The issue of deconfinement is more subtle, since it cannot be rigorously defined, but we estimate that confinement 
effect becomes of decreasing importance as we raise the temperature from the chiral temperature to that of the Hagedorn temperature.  
At zero temperature, we estimate the transition baryon chemical potential to be close to the nucleon mass \cite{Koch:2024qnz,McLerran:2024rvk}.
\begin{figure}
    \includegraphics[width=0.92\textwidth]{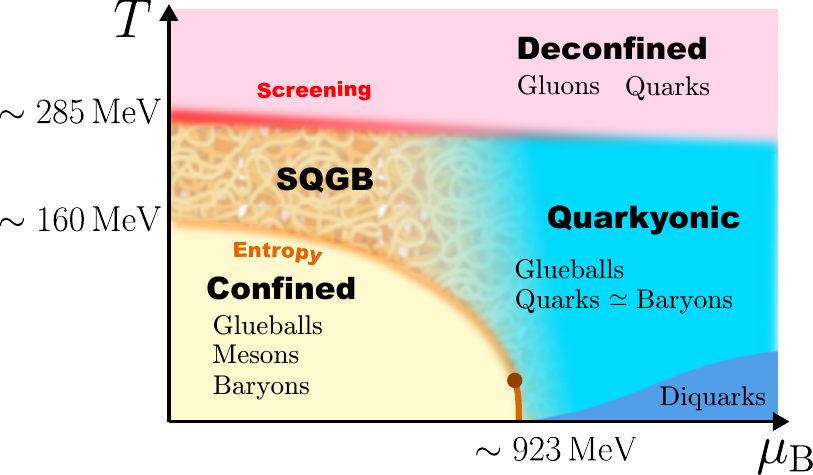}
    \caption{\footnotesize New and realistic phase diagram for $\Nc=3$ including a window of the SQGB regime. Figure is from Ref. \cite{Fujimoto:2025sxx}}
    \label{fig:diagram_N3}
\end{figure}
This is because we expect the  density  where this occurs to be not too much greater than that of nuclear matter, so that the nucleon 
Fermi momentum is $k_F << M_N$, and $k_F^2/2M^2 << 1$, hence $\mu_B$ is close to $M_N$.  In the high density 
Quarkyonic matter, various types of color superconducting phases can occur \cite{Rapp:1997zu,Alford:1998mk}.

 \subsection{Conclusions}

The generic features of Quarkyonic matter allow for a rapid transition from a soft nuclear matter equation of state to a hard 
Quarkyonic equation of state.  The Idylliq model provides an example of an explicitly solvable theory which has this property.

There are many issues\rr[,] theoretical and phenomenological\rr[,] which need to be addressed.  Clearly there are many possible models 
that might be constructed using different distribution functions for quarks inside of nucleons.  In the explicit model above, a singularity 
in $k$ exists and one must ask if this is an artifact of the model or is it possible to justify in some limit of QCD. 
 Moreover, the model probability distribution chosen is valid only for non-relativistic systems.   The sound velocity has an unphysical singularity.  How is this singularity tamed?

The Idylliq model is a free theory of nucleons.  What are the properties of this theory when one includes interactions of pions for  low densities?  Is this a sensible theory?

How does one generalize this theory to include the effects of finite temperature?  For describing the dynamics of neutron star collisions, one needs a theory 
at low but finite temperature and high baryon  density.

The equation of state dependence on isospin, and including the effect of hyperons is non-trivial due to the highly occupied phase space of quarks. 
Do hyperons strongly affect the equation of state computed in their absence?  Is the rapid rise in the sound velocity found in neutron stars also characteristic of zero isospin matter?

Can one determine the sound velocities dependence on density in nuclear matter at low temperature from comparison of computation 
to experiment for heavy ion collsions \cite{Sorensen:2021zme,Oliinychenko:2022uvy,Yao:2023yda}?

\section*{Acknowledgements}
L.M.\ thanks partial support from the Institute for Nuclear Theory which is funded in part by the INT's U.S.\ Department of Energy grant No.\ DE-FG02-00ER41132.

\bibliography{references.bib}

\end{document}